\documentclass [12 pt] {article}

\catcode`\@=11

\@addtoreset{equation}{section}
\catcode`\@=10

\def\half{\mbox{$\frac{1}{2}$}}
\def\dddot#1{\mathinner{\buildrel\vbox{\kern5pt\hbox{...}}\over{#1}}}
\def\ddddot#1{\mathinner{\buildrel\vbox{\kern5pt\hbox{....}}\over{#1}}}

\begin {document}

\begin {center}
{\Huge The Occurrence of a Triple $-1 $ Resonance in the Standard Singularity Analysis}\\[3 mm]
{\large K Andriopoulos \& PGL Leach\footnote {permanent address: School of Mathematical Sciences, University of KwaZulu-Natal, Private Bag X54001 Durban 4000, Republic of South Africa}\\[3 mm]
Department of Information and Communication Systems Engineering\\ University of the Aegean, Karlovassi 83 200, Greece}\\[3 mm]
\makeatletter email: kand@aegean.gr; leachp@ukzn.ac.za, leach@math.aegean.gr \makeatother
\end {center}

\noindent {\bf Abstract}: Any careful singularity analysis of the Mixmaster Universe uncovers the instance of a triple $-1$ resonance. The Mixmaster Universe does not exhibit a closed-form solution and so a correct interpretation of the meaning of the triple $-1$ resonance is difficult. We provide a system of differential equations for which both a closed-form solution and a triple $-1$ resonance exist.

\vspace*{3mm}


\vspace*{2mm}


\strut\hfill

\section {Introduction}

\strut\hfill

Our understanding of the workings of a physical system is generally enhanced by mathematical analysis provided that analysis is undertaken in a state of confidence in the meaning of the results obtained.  One way to develop such confidence is to have experience of the application of that analysis to a model system of known properties.  This is not a perfect route since in exploring the unknown one may stray so far from the known that prior experience cannot help.  In this Letter we intend to highlight this problem in terms of the singularity analysis of ordinary differential equations.  In Cosmology the Bianchi Type IX model, commonly known as the Mixmaster Universe, has received a number of investigations through singularity analysis\footnote {There have been other approaches and the reader is referred to \cite {Andriopoulos 07 a} for a recent summary of the literature.} \cite {Contopoulos 93 a, Contopoulos 94 a, Contopoulos 95 a, Cotsakis 94 a, Demaret 96 a, Latifi 94 a, Bountis 97 a, Springael 98 a, Andriopoulos 07 a}. The conclusions reached reveal an interesting evolution in the various interpretations of the results of the analysis.  The first paper \cite {Cotsakis 94 a}\footnote {Although the first of the three papers by Contopoulos and his coworkers appeared earlier in the Journal of Physics A: Mathematical and General, the paper by Cotsakis {\it et al} was submitted over a month earlier to the same journal, but it did not appear until some seven or eight months later.} reported a four-parameter particular solution.  Initially Contopoulos {\it et al} concluded that the system of differential equations describing the Mixmaster Universe satisfied the Painlev\'e criterion for integrability.  Subsequently Contopoulos {\it et al} moved to the position that the equations were not integrable due to the presence of essential singularities.  This opinion has been reinforced by the conclusions of Demaret and Scheen, Latifi {\it et al} and Springael {\it et al}.  More recently the present authors proposed that there existed two patterns of singularity behaviour, if one omits cyclic permutations, both of which possess the full number of constants of integration\footnote {Since the differential equations of the model are three of the second order, this means six.  As there is a constraint, the number is reduced to five.  Since the constraint is a particular value of a first integral of the system of differential equations, it plays no role in the analysis except to reduce the solution to existing on a surface in the space of initial conditions.} \cite {Andriopoulos 07 a}.

\strut\hfill

It is certainly not unreasonable to wonder why various investigations of the same system have led to a variety of interpretations.  There appear to be three sources for concern.  

\strut\hfill

The first is the understanding of the meaning of negative resonances -- apart from the generic ${-1} $ -- arising from the analysis.  In the case of an equation possessing only nonpositive resonances Lemmer {\it et al} \cite {Lemmer 93 a} had already demonstrated that one could have a Laurent expansion with the powers commencing at the singularity and descending to minus infinity.  This was formalised by Feix {\it et al} \cite {Feix 97 a}.  The interpretation of such a series is that it is valid on the exterior of a disc centred on the singularity so that it has the nature of an asymptotic expansion.

\strut\hfill

The second is concerned with the existence of both positive and negative resonances -- again apart from the generic ${- 1} $ -- for a given leading-order behaviour.  Latifi {\it et al} proposed that these represented two particular solutions, one descending and one ascending, since the number of constants of integration for each series would be less than the number required \cite {Latifi 94 a}.  In a recent paper \cite {Andriopoulos 06 a} it was demonstrated by way of explicit example that the case of resonances of mixed sign was simply a Laurent series valid on an annulus centred on the singularity.

\strut\hfill

The third source for concern is the appearance of a triple ${-1} $ resonance for one of the branches of the equations of the Mixmaster Universe.  Originally Cotsakis {\it et al} \cite {Cotsakis 94 a} thought that it indicated a four-parameter particular solution.  Then Contopoulos {\it et al} \cite {Contopoulos 93 a} proffered the opinion that it was due to the forcing of the three functions to have a singularity at the same point.  In our more careful analysis -- having benefited from the evolution of ideas over the past decade -- we found that the triple ${-1} $ resonance admitted three constants of integration \cite {Andriopoulos 07 a}.  Indeed the Laurent series contained the full set of constants of integration and would exist on an annulus since there were both positive and negative resonances.  There were two Laurent expansions.  One was moderately conventional and would exist on a punctured disc centred on the singularity and then there was the one on the annulus.  This result melds very nicely with the numerical studies of Bountis and Drossos \cite {Bountis 97 a}.  A matter of particular interest in this case is that the presence of the triple ${-1} $ resonance indicates a Laurent series of this type.  It is a commonplace that the existence of a resonance at ${-1} $ is generic and is related to the location of the movable singularity. That was contradicting the whole meaning of the resonances which were supposed to indicate the first entrance of yet another arbitrary constant in the expansion \cite{Dimas}. The reasoning strengthens when we move to systems of differential equations and that a multiple ${-1} $ resonance can indicate additional constants of integration whereas for a single equation it is not good news.

\strut\hfill

In this Letter it is our intention to discuss the singularity analysis of a system of three second-order nonlinear ordinary differential equations of moderately complex form which nevertheless can be integrated explicitly.  Thus we can see how the various principal branches relate to the general solution.

\strut\hfill

\section {The System and Its Solution}

\strut\hfill

We consider the system
\begin {eqnarray}
\ddot {x} +3\left (x\dot{x} + y\dot {z} + z\dot {y}\right) + x ^ 3 + y ^ 3 + z ^ 3 + 6xyz & = & 0\nonumber \\
\ddot {y} +3\left (x\dot {y} + y\dot {x} + z\dot {z}\right) + 3\left (x ^ 2y + y ^ 2z + z ^ 2x\right) & = & 0 \label {2.1} \\
\ddot {z} +3\left (x\dot {z} + y\dot {y} + z\dot {x}\right) + 3\left (x y ^ 2 + y z ^ 2 + z x ^ 2\right) & = & 0\nonumber
\end {eqnarray}
which has the three Lie point symmetries
\begin {equation}
\Gamma_1 = \partial _t,\quad \Gamma_2 = - t \partial _t + x \partial _x + y \partial _y + z \partial _z \,\,\,\mbox {\rm and}\,\,\, \Gamma_3 = - t^2 \partial _t + 2 (-1 + t x) \partial _x + 2 t y \partial _y + 2 t z \partial _z \label {2.4}
\end {equation}
with the algebra $sl (2,R) $.

\strut\hfill

The solution of the system (\ref {2.1}) is easily found to be
\begin {eqnarray}
y(t) & = & \displaystyle {\frac{C^2 \dot {A}+A^2 \dot {B}+B^2 \dot {C}-A B \dot {A}  -B C \dot {B} -A C \dot {C} }{A^3+B^3+C^3-3 AB C}}  \label {2.5} \\
z(t) & = & \displaystyle {\frac{B^2 \dot {A}+ C^2 \dot {B} +A^2 \dot {C} -A C \dot {A} -A B \dot {B}  -B C \dot {C} }{A^3+B^3+C^3-3 A BC}}, \nonumber 
\end {eqnarray}
where the functions $A (t) $, $B (t) $ and $C (t) $ are given by
\begin {eqnarray}
A (t) & = & a_0+ b_0t + c_0t ^ 2 \nonumber \\
B (t) & = & a_1+ b_1t + c_1t ^ 2 \label {2.6} \\
C (t) & = & a_2 + b_2t + c_2t ^ 2 \nonumber
\end {eqnarray}
and are introduced to render the expressions in (\ref {2.5}) to be of a manageable size.  Although there are nine arbitrary constants listed in (\ref {2.6}), only six in the solutions, (\ref {2.5}), are in fact independent.

\strut\hfill

It is evident from (\ref {2.5}) and (\ref {2.6}) that the solution of the system is analytic away from its polelike singularities.  Thus we can proceed with the singularity analysis in the sure knowledge that we can interpret the results in light of our possession of the explicit solution.  Such explicitly integrable systems may not give a complete answer to every question, but they do help to prevent the making of unjustified inferences.

\strut\hfill

\section {Singularity Analysis}

\strut\hfill

Since system (\ref {2.1}) possesses the self-similar symmetry, $\Gamma_{2} $, all terms are dominant\footnote {Here we do not make an investigation of subdominant behaviours since the analysis of the all-terms-dominant case provides more than sufficient results for our present purposes.} and it is obvious from the symmetry that the exponent of the leading-order terms is ${-1} $.  We make the usual substitution, $x = \alpha\tau ^ {-1} $, $y = \beta\tau ^ {-1} $ and $z = \gamma\tau ^ {-1}$, where $\tau=t-t_0$, into the equations.  The resulting set of algebraic equations is
\begin {eqnarray}
2 \alpha-3 \alpha^2-6 \beta \gamma+\alpha^3+\beta^3+\gamma^3+6 \alpha \beta \gamma & = & 0 \nonumber \\
2 \beta-6 \alpha \beta-3 \gamma^2+3 \alpha^2 \beta+3 \beta^2 \gamma+3 \alpha \gamma^2 & = & 0 \label {2.7} \\
2 \gamma-6 \alpha \gamma-3 \beta^2+3 \alpha \beta^2+3 \beta \gamma^2+3 \alpha^2 \gamma & = & 0 \nonumber
\end {eqnarray}
which provides a large number of solutions -- indeed $27$.  For each one of these  solutions which does not  contain a  zero value for the coefficient of a leading-order term -- that would make for a particular solution -- we can perform the calculations for the resonances.  We make the substitutions, $x = \alpha\tau ^ {-1} +\mu\tau ^ {s -1} $, $y = \beta\tau ^ {-1} +\nu\tau ^ {s -1} $ and $z = \gamma\tau ^ {-1} +\xi\tau ^ {s - 1} $,  where $s $ is the value of the resonance at which arbitrary constants enter the Laurent expansion, into the full equations since all terms are dominant for this leading-order exponent.  We summarise our results in Table I.

\strut\hfill

\begin{center}
TABLE I BE HERE

\vspace*{2mm}

{\bf Table I}: Coefficients of the leading-order terms and corresponding resonances for system (\ref {2.1}) in the case that all terms are dominant.
\end{center}

\strut\hfill

The order of presentation of the different possible patterns of values of resonances in Table I is dictated by the number of resonances of value ${-1} $ which happens to match an ordering of the values of $\alpha $ from zero to two by increments of one third.  Within each triplet we note that the sums of the values of $\beta $ or $\gamma $ in the second and third members of the triplet are the negatives of the value of the same for the first member of the triplet. Doubtless there could be other orderings based upon some other criteria.  For example one could take the sum of the values of $\alpha $, $\beta $ and $\gamma $ and use this to order the sets of the values of the resonances.  It is quite evident that this system of second-order equations displays an unexpected richness in both the coefficients of the leading-order terms and the values of the resonances.  In the case of certain sequences of scalar equations, such as the Riccati and Emden-Fowler sequences, this richness of property has been observed \cite {Euler 07 a, Andriopoulos 07 b, Leach 07 a} and, if one could construct a sequence commencing with the present system by finding its recursion operator, the patterns could become even more exotic.

\strut\hfill

In our discussion we ignore the sets of coefficients of leading-order behaviour which contain a zero or more.  

The first triplet gives a standard representation of a Right Painlev\'e Series in that the nongeneric resonances are positive\footnote {The ${- 1} $ resonance is always to be expected in this analysis \cite{Goriely}.  The interpretation of its meaning is another matter and is addressed below with this system providing a specific medium. See also \cite{Dimas}.}.  

The second triplet has one of the ${+ 1} $ resonances replaced by a ${- 2}$.  We now have a case of resonances of mixed sign\footnote {Since the ${- 1} $ resonance is expected, curiously it is not normally taken to be counted in a discussion of the existence of resonances of mixed sign.}.  

For the third triplet one of the ${+ 2}$ resonances of the first triplet has moved down to ${- 1} $ so that we have a double resonance at ${-1} $.  We return to this case below.  

The following two triplets are related to the third in the sense that one of the ${+ 1}$ resonances has become ${- 2}$.
  
In the instance of the sixth triplet a second of the ${+ 1} $ resonances has moved down to the value ${- 2} $.  

At the seventh triplet the remaining ${+ 2} $ resonance has made the third of the ${-1} $ resonances.  

When we come to the eighth triplet, the pattern of resonances is the direct opposite of that for the first triplet.

\strut\hfill

Corresponding to each value of the resonances for each of the sets of $\{\alpha,\beta,\gamma\} $ there is an eigenvector.  To present them for all the possible coefficients of the leading order terms would be pleonastic.  However, we believe that it is informative to consider the third and the eighth triplets.  In the case of the third triplet we have the intrusion of a double resonance at the so-called generic value ${-1} $.  For the (seventh and) eighth triplet we have a triple ${-1} $ resonance.  

\strut\hfill

For the first member of the third triplet the eigenvectors are (in the order corresponding to the ordering of the values of the resonances)
\begin {equation}
\left [\begin {array} {r}
1\\0\\-1\end {array}\right]\mu
+
\left [\begin {array} {r}
0\\1 \\-1\end {array}\right]\nu,\quad
\left [\begin {array} {r}
\mu\\ \nu\\ \xi \end {array}\right],\quad
\left [\begin {array} {r}
1\\1\\1\end {array}\right]\mu, \label {2.8}
\end {equation}
where $\mu $, $\nu $ and $\xi $ are parameters, and for the second
\begin {equation}
\left [\begin {array} {r}
1\\0\\\half (1 + i\sqrt {3})\end {array}\right]\mu
+
\left [\begin {array} {r}
0\\1 \\\half (1 -\sqrt {3})\end {array}\right]\nu,\quad
\left [\begin {array} {r}
\mu\\\nu\\\xi \end {array}\right],\quad
\left [\begin {array} {r}
-\half (1 +i\sqrt {3})\\1\\-\half (1 -i\sqrt {3})\end {array}\right]\mu. \label {2.9}
\end {equation} 
The eigenvectors for the third member of the triplet have $i $ replaced by $-i $, {\it ie} they are as if the complex conjugate was taken with the parameter(s) being regarded as real.

\strut\hfill

Curiously the eigenvectors for the eighth triplet are precisely the same as for the third triplet.

\strut\hfill

The importance lies in the fact that for each of the combinations of exponents of the leading-order terms and their resonances the number of arbitrary constants is the correct number.

\strut\hfill

\section {Interpretation}

\strut\hfill

The first triplet presents Right Painlev\'e Series as the solution.  In addition to the arbitrary location of the singularity three constants enter at the resonance ${+ 1} $ and two at the resonance ${+ 2} $.  The first few terms of the Laurent expansion for the first member of the triplet are
\begin {eqnarray*}
x & = & \mbox{$\frac{1}{3}$}\tau^{-1}  + a_0 - \left\{b_1+c_1+ (a_0+b_0+c_0) ^2\right\}\tau + \half\left\{(a_0 -b_0) ^3+ (c_0 - a_0) ^ 3 -3 a_0 ^ 2 (b_0+c_0) \right. \\
&& \left. \phantom{\half} + 3 (b_1+c_1) (a_0 -b_0\right\}\tau ^2+ \ldots \\
y & = & \mbox{$\frac{1}{3}$}\tau^{-1}   + b_0 + b_1\tau - \mbox {$\frac {3} {2} $}\left\{c_0c_1 +c_0^ 2 a_0 -b_0c_0^ 2 - c_1b_0 -2 a_0b_0c_0 -b_0^ 2c_0 -2 a_0b_0^2 -b_0 ^3+ a_0b_1 -b_0b_1\right\}\tau ^2 + \ldots \\
z & = & \mbox{$\frac{1}{3}$}\tau^{-1} + c_0 + c_1\tau + \mbox {$\frac {3} {2} $}\left\{c_0 ^ 3 + 2 a_0c_0 ^ 2 +b_0c_0 ^ 2 + 2 a_0b_0c_0+b_0 ^ 2c_0 - a_0b_0 ^ 2+c_0 (b_1+c_1) - a_0c_1 - b_0b_1\right\}\tau ^2 + \ldots.
\end {eqnarray*}
It is easily appreciated that the coefficients become unmanageably complicated within a few terms.  Nevertheless it is evident that three arbitrary constants, $(a_0,\, b_0,\, c_0)$, enter at the first resonance and two arbitrary constants, $(b_1,\,c_1)$, at the second resonance.  Moreover there is consistency at the second resonance and the arbitrary constants introduced at the first resonance remain arbitrary.  Altogether the first triplet belongs to the so-called usual case of the singularity analysis with the single generic resonance at ${-1} $ and multiple positive resonances with the fortunate property that their geometric multiplicity equals their algebraic multiplicity at each resonance.

\strut\hfill

All the other possible patterns for the resonances are of mixed sign and give rise to Laurent expansions in which the exponent goes to plus and minus infinity and so each represents an expansion on an annulus centred on the singularity \cite {Andriopoulos 06 a}.  This is to be expected when there is a resonance at ${-2} $.  However, it is not so expected in the third triplet where one only has a double `generic' ${- 1} $ resonance.  Unfortunately there does not appear to exist an algorithmic procedure which enables one to calculate the coefficients of the series by substituting them into system (\ref {2.1}).  Nevertheless we can see how two arbitrary constants enter at the resonance ${- 1} $ by partially following the example of Latifi {\it et al} \cite {Latifi 94 a} to construct a particular solution by putting to zero the constants of integration which enter at some of the resonances.  As we are interested only in the descending series, we set to zero the three arbitrary constants occurring at the resonance ${+1} $ and one arbitrary constant entering at ${+ 2} $, {\it ie}, we write the solutions in the form of Left Painlev\'e Series.  We choose the first member of the triplet to illustrate our point.  We obtain
\begin {eqnarray*}
x & = & \displaystyle {\frac {2} {3\tau} + \frac {a_2} {\tau ^ 2} +\frac {a_2 - 2a_2b_2 - 2b_2 ^2} {\tau ^ 3} - \frac {9 (a_2 + b_2) a_2b_2} {\tau ^ 4}} + \ldots \\
y & = & -\displaystyle {\frac {1} {3\tau} + \frac {b_2} {\tau ^2} +\frac {a_2 + 4a_2b_2 + b_2 ^2} {\tau ^ 3} + \frac {3 (a_2 ^3+ 3 a_2 ^2b_2 -b_2 ^ 3)} {\tau ^ 4}} + \ldots \\
z & = & -\displaystyle {\frac {1} {3\tau}  - \frac {a_2 +b_2} {\tau ^2} -\frac {a_2 +a_2b_2 - b_2 ^2} {\tau ^ 3} - \frac{3 (a_2 ^ 3 -3 a_2b_2 ^2 -b_2 ^ 3)} {\tau ^ 4}} + \ldots. 
\end {eqnarray*}
We see that the expansion is in terms of two arbitrary constants, $a_{2} $ and $b_{2} $.  It might be thought that there are three arbitrary constants since we also have the location of the movable singularity.  However, if we expand the first term in each series in the usual way to explain the existence of the generic resonance, we see that $t_0$ is absorbed into $a_2$ and $b_2$ \cite {Dimas}.  We emphasise that we took the particular values of the constants of integration made above so that we could provide an accurate representation of the solutions and the way in which the two arbitrary constants entered into the series at the double ${-1} $ resonance.  The general solution must have the arbitrary constants from the positive resonances as well as the two entering at the negative resonance.

\strut\hfill

In (\ref {2.8}), which we recall is common to both the first member of the third triplet and the first member of the eighth triplet, the eigenvector corresponding to the triple ${-1} $ resonance is
\begin {equation}
\bf {v} = \left (\begin {array} {c} 1\\0\\0 \end {array}\right)\mu
+ \left (\begin {array} {c} 0\\1\\0 \end {array}\right)\nu
+ \left (\begin {array} {c} 0\\0\\1 \end {array}\right)\xi, \label {3.1}
\end {equation}
{\it ie}, there are three arbitrary constants entering at this resonance.  As in the case of the first triplet the location of the movable singularity is absorbed into these three constants so that system (\ref {2.1}) has six arbitrary constants with the other three entering at the double ${-2} $ and single ${+1} $ resonances.

\strut\hfill

\section {Conclusion}

\strut\hfill

We have seen by way of explicit example on the system, (\ref {2.1}), of three nonlinear second-order ordinary differential equations that the so-called generic resonance at ${-1} $ plays a more complex role in the singularity analysis than it is generally accorded in the literature.  The reasons for this to occur are unknown and perhaps it would be better not to speculate.  Speaking of our own experiences our understanding of the singularity analysis has evolved in time from the examination of many explicit examples.  In retrospect one wonders how one could have failed to appreciate that the essential feature of the singularity analysis is the determination of Laurent expansions as solutions to the differential equations under consideration.  Then one may have recalled that the expansion of a function as a Laurent series can be at various parts of the complex plane.  It can be in the vicinity of the singularity, which is where we have a Right Painlev\'e Series.  It could be beyond the last singularity and so have the nature of an asymptotic expansion which is where we have a Left Painlev\'e Series.  Finally it could be between two singularities separated from the singularity about which the expansion is being made.  This gives a series in which the exponent can range from minus infinity to plus infinity.

\strut\hfill

It is with the appreciation of the different domains of convergence of the results of one's analysis that one understands the three possible types of series.  That one obtains a series differing in appearance \cite{Baldwin 06 a} from what the `traditional' analysis allows does not mean that one has, for example, an essential singularity.  There may be an essential singularity, but the evidence provided by the standard singularity analysis is not going to enable one to state that as a certainty as was the case for the Mixmaster Universe \cite{Contopoulos 95 a}.

\strut\hfill

The value of system (\ref {2.1}) is that we have the solutions in closed form.  Consequently we can look at the results of the analysis and see whether they say anything peculiar about an integrable system.  Moreover, when we obtain similar results from a system the integrability of which is unknown, we must be careful not to make an inference which is unsupported by the evidence.

\strut\hfill

Finally we have seen that a multiple resonance at ${-1} $ can indicate the entry of more than one constant into the solution.  This is important in a consideration of integrability.  Perhaps an unexpected result was that a multiple resonance at ${-1} $ without any other negative resonances could presage Laurent series going to minus infinity.

\strut\hfill

There is a question which remains unanswered.  The Laurent series which arise as a result of the singularity analysis can be expected to have limitations on the regions in which they are convergent.  If one has the three possible types of series, one would expect that the whole of the complex plane would be covered apart from the singularities.  What happens, as we have in the case of (\ref {2.1}), when the series are convergent on the interior of a punctured disk and the interior of an annulus\footnote{In fact system (\ref{2.1}) has six singularities. According to the analysis these singularities create one punctured disc and five annuli.} centred on a singularity?  There are no asymptotic series (Left Painlev\'e Series).  Is it necessary to cover the whole complex plane, apart from singularities, to be able to infer integrability?

\strut\hfill

\section*{Acknowledgements}

\strut\hfill

PGLL thanks DICSE, University of the Aegean, for the provision of facilities during the period when this work was undertaken and the University of KwaZulu-Natal for its continued support. KA thanks the State (Hellenic) Scholarship Foundation and the University of KwaZulu-Natal. 

\strut\hfill

\newpage

\[
\begin {array}{|c|c|c|}\hline
&&\\
{\bf Coefficients \,\, \{\alpha,\,\beta,\,\gamma\} }&\mbox{\rm {\bf Resonances}} & \, \mbox{\rm {\bf Triplet}} \, \\
&&\\
 \hline
\{0, 0, 0\} &&\\ &&\\
\left\{ \frac{1}{3}, \frac{1}{3}, \frac{1}{3}\right\}& -1,\, 1 (3),\, 2 (2) &\\
\left\{ \frac{1}{3}, \frac{1}{6}\left(-1-i \sqrt{3}\right), \frac{1}{6} \left(-1+i \sqrt{3}\right)\right\}& -1,\, 1(3),\, 2(2) & {\bf 1}\\
\left\{ \frac{1}{3}, \frac{1}{6} \left(-1+i \sqrt{3}\right), \frac{1}{6} \left(-1-i \sqrt{3}\right)\right\}& -1,\, 1(3),\, 2(2) &\\ &&\\
\left\{ \frac{2}{3}, \frac{2}{3}, \frac{2}{3}\right\}& -2,\, -1,\, 1 (2),\, 2 (2) &\\
\left\{ \frac{2}{3}, \frac{1}{3} \left(-1-i \sqrt{3}\right), \frac{1}{3} \left(-1+i \sqrt{3}\right)\right\}& -2,\, -1,\, 1(2),\, 2(2) & {\bf 2}\\
\left\{ \frac{2}{3}, \frac{1}{3}\left(-1+i \sqrt{3}\right), \frac{1}{3} \left(-1-i \sqrt{3}\right)\right\}& -2,\, -1,\, 1(2),\, 2(2) &\\ &&\\
\left\{ \frac{2}{3}, -\frac{1}{3}, -\frac{1}{3}\right\}& -1 (2),\, 1 (3),\, 2 &\\
\left\{ \frac{2}{3}, \frac{1}{6} \left(1-i \sqrt{3}\right), \frac{1}{6} \left(1+i \sqrt{3}\right)\right\}& -1(2),\, 1(3),\, 2 & {\bf 3}\\
\left\{ \frac{2}{3}, \frac{1}{6} \left(1+i \sqrt{3}\right), \frac{1}{6} \left(1-i \sqrt{3}\right)\right\}& -1(2),\, 1(3),\, 2 &\\ &&\\
\{ 1, 0, 0\}&&\\ &&\\
\left\{ 1, \frac{i}{\sqrt{3}}, -\frac{i}{\sqrt{3}}\right\}& -2,\, -1(2),\, 1(2),\, 2 &\\
\left\{ 1, \frac{1}{6} \left(-3-i \sqrt{3}\right), \frac{1}{6} \left(-3+i\sqrt{3}\right)\right\}& -2,\, -1(2),\, 1(2),\, 2 & {\bf 4}\\ 
\left\{ 1, \frac{1}{6} \left(3-i \sqrt{3}\right), \frac{1}{6} \left(3+i \sqrt{3}\right)\right\}& -2,\, -1(2),\, 1(2),\, 2 & \\ &&\\
\left\{ 1, -\frac{i}{\sqrt{3}}, \frac{i}{\sqrt{3}}\right\}& -2,\, -1(2),\, 1(2),\, 2 &\\
\left\{ 1, \frac{1}{6} \left(-3+i \sqrt{3}\right), \frac{1}{6} \left(-3-i \sqrt{3}\right)\right\}& -2,\, -1(2),\, 1(2),\, 2 & {\bf 5}\\
\left\{ 1, \frac{1}{6} \left(3+i \sqrt{3}\right), \frac{1}{6} \left(3-i \sqrt{3}\right)\right\}& -2,\, -1(2),\, 1(2),\, 2 & \\ &&\\
\left\{ \frac{4}{3}, -\frac{2}{3}, -\frac{2}{3}\right\}& -2 (2),\, -1 (2),\, 1,\, 2 &\\
\left\{ \frac{4}{3}, \frac{1}{3} \left(1-i \sqrt{3}\right), \frac{1}{3} \left(1+i \sqrt{3}\right)\right\}& -2 (2),\, -1 (2),\, 1,\, 2  & {\bf 6}\\
\left\{ \frac{4}{3}, \frac{1}{3} \left(1+i \sqrt{3}\right), \frac{1}{3} \left(1-i \sqrt{3}\right)\right\} & -2 (2),\, -1 (2),\, 1,\, 2  &\\ &&\\
\left\{ \frac{4}{3}, \frac{1}{3}, \frac{1}{3}\right\}&  -2,\, -1 (3),\, 1 (2) &\\
\left\{ \frac{4}{3}, \frac{1}{6} \left(-1-i \sqrt{3}\right), \frac{1}{6} \left(-1+i \sqrt{3}\right)\right\}& -2,\, -1(3),\, 1(2) & {\bf 7}\\
\left\{ \frac{4}{3}, \frac{1}{6}\left(-1+i \sqrt{3}\right), \frac{1}{6} \left(-1-i \sqrt{3}\right)\right\} & -2,\, -1(3),\, 1(2)  &\\ &&\\
\left\{ \frac{5}{3}, -\frac{1}{3}, -\frac{1}{3}\right\} & -2 (2),\, -1 (3),\, 1 &\\
\left\{ \frac{5}{3}, \frac{1}{6}\left(1-i \sqrt{3}\right), \frac{1}{6} \left(1+i \sqrt{3}\right)\right\} & -2 (2),\, -1 (3),\, 1  & {\bf 8} \\
\left\{ \frac{5}{3}, \frac{1}{6} \left(1+i \sqrt{3}\right), \frac{1}{6} \left(1-i \sqrt{3}\right)\right\} & -2 (2),\, -1 (3),\, 1 & \\ &&\\
\{ 2,0,0\} && \\
\hline
\end{array}\]

\end {document}